\documentstyle[12pt,epsfig]{article}

\begin{document}
\topmargin -1.4cm
\oddsidemargin -0.8cm
\evensidemargin -0.8cm 

\title{Estimate of a non-helical electroweak 
primordial\\ bootstrap field today}

\vspace{1.5cm}

\author{Poul Olesen\\
{\it  The Niels Bohr Institute}\\
{\it Blegdamsvej 17, Copenhagen \O, Denmark }}

\maketitle

\begin{abstract}
We estimate the magnitude today of the primordial magnetic field 
originating at the
electroweak phase transition. We find that the field,
which at the electroweak phase transition is originally of order 
$10^{23}-10^{24}$ G correlated over the Hubble scale, today is  of
order $10^{-14}$ G at a scale of order 2 kpc. This result is consistent
with the lower limit on the strength of intergalactic magnetic fields 
obtained by Neronov and Vovk \cite{neronov} from observations of TeV blazars.
The field is non-helical. We compare our results with the helical case
discussed by Field and Carroll. 
\end{abstract}

\thispagestyle{empty}

\vskip0.5cm

\section{Introduction}

It is  well known that the standard electroweak theory is in agreement
with experiments. It is therefore of interest to investigate cosmological
consequences of this theory and compare these to observations.

In this note we therefore estimate the cosmological evolution  of the 
bootstrap magnetic field generated at the
electroweak phase transition \cite{pol} and compare our results to
some observations. This field is originaly of
order $10^{23}-10^{24}$ G over the Hubble scale. During the evolution of the
universe it reduces to a field of order $10^{-14}$ G today. This magnitude is
consistent with the lower limit $B\geq 3\times 10^{-16}$ G obtained by
Neronov and Vovk \cite{neronov} from observations of TeV blazars. This
bound depends on the magnetic correlation length $\lambda$ and is
lifted as $\lambda^{-1/2}$ if $\lambda$ is much smaller than of the order
a megaparsec. In our case the correlation length is of order 2 kpc,
which corresponds to a lower limit of order $\sim 10^{-15}$. Being a lower limit 
this is perfectly consistent with our result $2.55\times 10^{-14}$. 

It should be mentioned that long time ago Vachaspati found the first
electroweak candidate
for a primordial magnetic field \cite{tand}. His solution differs fundamentally
from ours.

The contents of this note is the following:

In Section 2 we discuss the electroweak  solution originaly
due to Ambj\o rn and the author \cite{amb1}-\cite{amb3}. In section 3 we 
introduce the self-similarity found by Zrake \cite{zrake} and the
author \cite{poul} and discuss how this relation can be used to estimate the
magnetic field and the correlation length today of the bootstrap 
field originating at the electroweak phase transition. In section 4
we compare with the helical case, in partilar with the maximally
helical case discussed in detail by Field and Carroll \cite{field}.
Section 5 cocludes the note.

\section{The electroweak bootstrap solution}

The magnetic bootstrap solution of the electrowak theory is simplest in 
the Bogomolnyi case where the Higgs mass equals the $Z$ mass. We
shall therefore mainly exhibit this special case. 
The standard electroweak  static energy can be written \cite{amb1}
\begin{equation}
{\cal E }={\cal E}_1+{\cal E}_2+{\cal E}_3+{\cal E}_4+{\cal E}_5-
\frac{g^2~\phi_0^4}{8\sin^2\theta}
+\frac{g~\phi_0^2}{2\sin\theta}B-\frac{g~\phi_0^2}{2\cos\theta}Z-
\frac{g}{2\cos\theta}\partial_j(\epsilon_{ij}Z_j\phi^2),
\label{energy}
\end{equation}
where
\begin{equation}
{\cal E}_1=|(D_1+iD_2)W |^2,
\end{equation}
\begin{equation}
{\cal E}_2=\frac{1}{2}\left(B-\frac{g}{2\sin\theta}~\phi_0^2-
2g\sin\theta~|W|^2\right)^2,
\end{equation}
\begin{equation}
{\cal E}_3=\frac{1}{2}\left(Z-\frac{g}{2\cos\theta}(\phi^2-\phi_0^2)-
2g\cos\theta~|W|^2\right)^2,
\end{equation}
\begin{equation}
{\cal E}_4=\left(\frac{g\phi}{2\cos\theta}Z_i+
\epsilon_{ij}\partial_j\phi\right)^2,
\end{equation}
and
\begin{equation}
{\cal E}_5=\left(\lambda-\frac{g²}{8\cos^2\theta}\right)~(\phi^2-\phi_0^2)².
\end{equation}
Here we took $W_2$=$iW_1\equiv iW$, $B$ is the magnetic field and $Z$ is the
corresponding $Z-$field. Also
\begin{equation}
D_i=\partial_i-ig(A_i\sin\theta+Z_i\cos\theta),
\end{equation}
where $A_i$ are the potentials for $B$ and $Z_i$ are the corresponding
potentials for the $Z$ field.

The Bogomolnyi method then consists in taking first ${\cal E}_5=0$, which
implies unrealistically that the Higgs mass equals the $Z$ mass. Then 
we minimize the energy
by taking the positive  terms ${\cal E}_i, i=1,...,4 $ to vanish. We refer 
to \cite{amb1} for a
discussion of  all these equations of motion. Here we just 
mention ${\cal E}_2=0$, i.e.
\begin{equation}
B=\frac{m_W^2}{e}+2e~|W|^2.
\label{ao}
\end{equation}
We mention that if the Higgs mass is not equal to the $Z$ mass there will be
higher order terms in powers of $|W|$ on the right hand side. 

The solution arizing from ${\cal E}_i=0, i=1,...,4$ is a bootstrap solution 
from the following point of view:
\vskip0.5cm

1)The magnetic field is generated like in a solenoid by a current 
consisting of charged $W$
vector meson fields. The action of thes fields is anti-screening
\cite{amb1}-\cite{amb3}.
\vskip0.3cm

2)The existence of the charged vector bosons is due to the magnetic
field, which is unstable \cite{nkn} 
above the threshold $m_W^2/e$ unless the $W$ mesons are present.
\vskip0.5cm

Eq.(\ref{ao}) implies that the magnetic field exceeds the threshold 
\begin{equation}
B\geq \frac{m_W^2}{e}\approx 3\times 10^{23}~{\rm G}.
\label{lower}
\end{equation}
It can be shown \cite{amb2} that in the realistic case  where the
Higgs mass is not equal to the $Z$ mass the upper limit including
contributions from the $|W|^2-$term is 
\begin{equation}
B\leq \frac{m_H^2}{e}\approx 7.64\times 10^{23}~{\rm G},
\label{upper}
\end{equation}
where $m_H$ is the Higgs mass. Compared to the total energy density at the
electroweak (EW) phase transition\footnote{We use \cite{field} 
that $\rho(t_{\rm EW}) \approx 1.6\times 10^{49}~{\rm G}^2$.}
 this gives
\begin{equation}
1.82\times 10^{-2}\geq \Omega_B^{\rm EW}\geq 2.8\times 10^{-3}
\label{omega}
\end{equation}
As discussed in the literature the solution of the equation of motion 
associated with Eq. (\ref{ao}) is a lattice 
of vortices  \cite{amb1}. This has been shown numerically \cite{amb3} as well as
mathematically \cite{math1}-\cite{math5}. This solution represents a
collective phenomenon in the sense that Eq. (\ref{ao}) does not have a
{\it single} vortex solution, only a lattice of an infinite number of vortices
as solution. Some other properties of Eq. (\ref{ao}) have recently been 
discussed in refs. \cite{manton1}-\cite{manton3}.

The lattice of flux tubes have a simple expression for the energy in the
Bogomolnyi case. Integrating over one lattice cell the expression (\ref{energy})
gives by use of periodicity \cite{amb1}
\begin{equation}
{\cal E}_{\rm one~cell}=\frac{m_W^2}{e}\int_{\rm one~cell}B~d^2x-\frac{m_W^4}{2e²}A
=2\pi \frac{m_W^2}{e^2}-\frac{m_W^4}{2e^2}A,~~\int_{\rm one~cell}B~d^2x=
\frac{2\pi}{e},
\end{equation}
where $A$ is the area of a cell. Here we used that the flux is 
quantized \cite{amb1}. Realistically, with the right Higgs mass, the energy 
becomes more complicated, with quadratic terms in the fields.

At the electroweak phase transition there is a high temperature which will cause
the flux tubes to oscillate. This is known also to happen in high 
temperature superconductors
where it has been shown that the Lindemann ratio\footnote{The
Lindemann ratio is the ration of the fluctuations of the tubes (or 
molecules) relative to the equilibrium distance. If the ratio exceeds
1/10 a liquid is formed.} is so large that a liquid state is formed, where the
flux tubes enter a (boiling) spaghetti state \cite{hightemp}. So we assume that
a similar phenomenon happens in our case. For us the main point is
that the high temperature state is statistically isotropic. We shall use this
to compute the subsequent evolution of the magnetic field to obtain its
present value. 

\section{Evolution of the magnetic field\\ in the expanding universe}

We now assume statistical isotropy and the validity of the standard
magnetohydrodynamics (MHD) in the form
\begin{equation}
\partial_t{\bf v}+{\bf (v\nabla) v}=-\nabla p+{\bf (\nabla\times B)\times B }
+\nu \nabla^2 {\bf v},~~\partial_t{\bf B}={\bf \nabla\times (v\times B)}+
\eta \nabla^2 {\bf B},
\label{mhd}
\end{equation}
where $\nu$ and $\eta$ are time independent constants. From isotropy and 
scaling invariance of the MHD equations it follows \cite{poul} that the energy
density
\begin{equation}
{\cal E}(k,t)=\frac{2\pi k^2}{(2\pi)^3}~\int d^3y~e^{i{\bf ky}}~
<B({\bf x},t)B({\bf x+y},t)>
\end{equation}
satisfies
\begin{equation}
{\cal E}(k,t)=\sqrt{\frac{t_0}{t}}~{\cal E}\left(k\sqrt{\frac{t}{t_0}},
t_0\right).
\end{equation}
We also define a mean field $B_{\rm rms}$ by
\begin{equation}
\frac{1}{2}~B_{\rm rms}^2\equiv \frac{1}{2}~<{\bf B(x},t)^2>=\int_0^\infty 
dk~{\cal E}(k,t).
\end{equation}
We next consider the evolution of the universe with the flat 
expanding metric
\begin{equation}
d\tau^2=dt^2-a(t)^2d{\bf x}^2=a(t)^2[(dt^*)^2-d{\bf x}^2]
\end{equation}
where $t^*$ is the conformal time, i.e.
\begin{equation}
t^*=\int \frac{dt}{a(t)},
\end{equation}
where $t$ is the Hubble time. The scaling of the energy density now
reads
\begin{equation}
{\cal E}(k,t_{\rm today})=\frac{a(t_{\rm EW})^4}{a(t_{\rm today})^4}\sqrt{\frac
{t_{\rm EW}^*}{t_{\rm today}^*}}~{\cal E}\left(k\sqrt{\frac{t_{\rm today}^*}
{t_{\rm EW}^*}},t_{\rm EW}\right).
\label{scaling}
\end{equation}
This was derived from isotropy and validity of the MHD equations. It may be 
that these assumptions are only valid slightly after the electroweak 
transition happening at the time
\begin{equation} 
t_{\rm EW}\approx 6\times 10^{-12}~{\rm sec}.
\end{equation}
In the following we shall assume that this difference is so small that it can
be ignored.

The magnetic coherence length $\lambda(t)$  is obtained by integrating over 
the energy,
\begin{equation}
\lambda(t)=<1/k>,~~~<1/k>=\int_0^\infty  {\cal E}(k,t)\frac{dk}{k}
\left/\int_0^\infty 
{\cal E}(k,t)dk \right .  .
\end{equation}
From the scaling (\ref{scaling}) we obtain
\begin{equation}
\lambda (t_{\rm today})=\frac{a(t_{\rm today})}{a(t_{\rm EW})}~
\sqrt{\frac{t_{\rm today}^*}{t_{\rm EW}^*}}~\lambda (t_{\rm EW}).
\label{coherence}
\end{equation}
Similarly we have from the scaling (\ref{scaling})
\begin{equation}
B_{\rm rms}(t_{\rm today})=\frac{a(t_{\rm EW})^2}{a(t_{\rm today})^2}~
\sqrt{\frac{t_{\rm EW}^*}{t_{\rm today}^*}}~B_{\rm rms}(t_{\rm EW}).
\label{rms}
\end{equation}
It should be noted that from Eq. (\ref{coherence}) there is an
amplification of the coherence length in addition to the general 
relativity scale factor $a(t)$
\begin{equation}
{\rm amplification ~factor}=\sqrt{\frac{t_{\rm today}^*}{t_{\rm EW}^*}}
\approx 5.7\times 10^6.
\label{amplification}
\end{equation}
Therefore, in terms of the Hubble time $t$ the amplification behaves as either
$t^{1/4}$ at early times before matter domination at $t_{\rm EQ}$,
\begin{equation}
t_{\rm EQ}=21.6\times 10^{22}~t_{\rm EW},
\end{equation}
or it behaves at at later times like $t^{1/6}$. 

Numerically we obtain
from Eq. (\ref{coherence})
\begin{equation}
\lambda (t_{\rm today}=13.8~{\rm Gyr})\approx 1.8~{\rm kpc},
\label{33}
\end{equation}
where we used that the initial coherence length of the field is given by
the Hubble radius,
\begin{equation}
\lambda(t_{\rm EW})=1/H_{\rm EW}\approx 0.4~{\rm cm}
\end{equation}
since the initial solution is not limited in space.

From Eqs. (9), (10), and (\ref{rms}) we have for the rms field
\begin{equation}
B_{\rm rms}(13.8~{\rm Gyr})\approx (1~{\rm to}~2.55)\times 10^{-14}~{\rm G}.
\label{bbound}
\end{equation}
These results are consistent with the lower limits obtained in 
\cite{neronov}, according to which the field should exceed $3\times 10^{-16}$G
with this bound improving as $\lambda^{-1/2}$ for correlation lengths
much smaller than a megaparsec. With our correlation length
of order 2 kpc the bound (\ref{bbound}) is thus rather close (within a
factor $\sim 10$)  to
the bound of Neronov and Vovk.

The rms field and the coherence length are statistical quantities. A more
informative quantity is the energy spectrum. The scaling relation 
(\ref{scaling}) relates the energy densities at different times. So
measuring the energy today, one can obain the spectrum at earlier times.
For example,
\begin{equation}
{\cal E}(k,t_{\rm EW}\approx 6\times 10^{-12}~{\rm sec})\approx
14.45\times 10^{67}\times{\cal E}(1.76\times k\times 10^{-7},
t_{\rm today}\approx 13.8~{\rm Gyr}),
\end{equation} 
allows knowledge of the spectrum at the electroweak phase transition.
The self-similarity scaling essentially says that the energy spectrum
does not change except kinematically.

There is a subtility in using Eq. (\ref{scaling}), since it assumes the validity
of the MHD equations at all times. These may not be valid exactly at the 
time of the
phase transition, but only slightly later when the creation of the field is
finished \cite{italien}. Therefore the time $t_{\rm EW}$ is in general only
close to the transition time. Numerical examples show that this effect is quite
small. For exampel, in the numerical calculations by Zrake \cite{zrake} after 
a fraction of an Alfv\'en time the magnetic energy spectrum relaxes to
a  self-similar power behavior.

\section{Comparison with a helical case}

The  bootstrap solution we have considered is non-helical, with ${\bf AB}$=0.
Some time ago Field and Carroll \cite{field} estimated how a maximally
helical field develops during the evolution of the universe. They assumed
an initial behavior at the electroweak phase transition given by
\begin{equation}
\lambda (t_{\rm EW})=f_\lambda/H_{\rm EW}\approx 0.4f_\lambda~{\rm cm}~~{\rm and}
~~B_{\rm rms}(t_{\rm EW})\approx 2\times 10^{25}f_B~{\rm G}.
\end{equation}
After maximally helical amplification Field and Carroll obtain
\begin{equation}
\lambda (t_{\rm today})\approx 20 f_\lambda~{\rm kpc}~~{\rm and}~~
B_{\rm rms}(t_{\rm today})\approx 4\times 10^{-10}f_B~ {\rm G },~~
{\bf AB}\neq 0.
\end{equation}
This should be compared to our results from the last section given in Eqs.
(\ref{33}) and (\ref{bbound}), i.e.
\begin{equation}
\lambda (t_{\rm today})\approx 1.8~{\rm kpc}~~{\rm and}~~
B_{\rm rms}(t_{\rm today})\leq 2.55\times 10^{-14}~ {\rm G },~~
{\bf AB}=0.
\end{equation}
We see that the coherence lengths differ  by a factor $\sim$ 10 if 
$f_\lambda =1$,
and the rms fields differ   by a factor $\sim 10^4$ if $ f_B=1$. The
latter case means that all energy at the electroweak phase transition
is magnetic. In
our case this is far from the case, since from Eq. (\ref{omega}) we see that
$\Omega^{\rm EW}_B\sim 3\times 10^{-3}$. If we limit ourselves to such an
$\Omega_B$ then the helical case would be reduced   by a factor $\geq\sim$10. 

In this
connection it should be mentioned that from a dimensional argument it is 
difficult to get the entire energy at the electroweak phase transition to
be magnetic, since (in natural units) the magnetic field has dimension
(mass)$^2$, so except if some large numerical factor is present it is
impossible to get a result for the magnetic field which is much larger
than what we obtained. There are simply not large enough masses to
produce an $\Omega_B\sim 1$ in the conventional 
electroweak theory! 

The results mentioned above shows that it is not necessary to have helicity
in order to obtain a type of inverse cascade. However, the enhancement
effect is larger with helicity present. For completeness, we compare the 
amplification factors in the two cases, namely for the helical
case \cite{field}
\begin{equation}
{\rm helical~amplification}=(t_{\rm EQ}/t_{\rm EW})^{1/3}=6\times 10^7~~
{\bf for}~~{\bf AB}\neq 0,
\end{equation}
and Eq. (\ref{amplification}) for the non-helical case with the
amplification factor
\begin{equation}
 \sim 6\times 10^6 ~{\rm for} ~{\bf AB}=0.
\end{equation}
Thus the difference in the amplification is a factor of 10.

Since the helical case gives a larger effect there has been many 
studies of the effects of the chiral anomaly on the evolution of
the (hyper-)magnetic field,  
see refs. \cite{10}-\cite{16}. In the modified MHD equations found
in some of these papers there is no simple scale invariance which allows
one to derive  a self-similar expression for the time development of the 
energy. However, when a field generated early "passes'' the 
electroweak phase transition there may be the possibility that this
helical field  have a further development similar to the magnetic properties
discussed in this note, in paticular if the usual MHD equations take
over from the modified ones at the transition \cite{11},\cite{15}.

We mention that Jedamzik and Sigl \cite{karsten} have shown, using
different methods, that appreciable primordial fields originating from
phase transitions is possible, even when viscosity and dissipation
is considered. These results as well as the estimate of  the
rms magnetic field today is in reasonable agreement with our result.

\section{Conclusion and discussion}

Primordial magnetic fields are generated at the electroweak phase transition. 
They could act as extremely weak extragalactic seed fields for various 
astrophysical dynamo effects. It is important for our result 
that the coherence length
is amplified by a factor of order $10^6$, as mentioned in Eq. 
(\ref{amplification}). If this had not been the case the coherence
length would have been of the order $\sim 10^{-3 }$ pc, which
would have been without interest cosmologically.

If the magnetic field discussed here is relevant as a cosmological seed
field it is of interest from the point of view of principles
that then this primordial field is related to the
non-Abelian vacuum, which is magnetic in nature. The flux tubes also exist above
the critical temperature, but the string (flux tube) tension vanishes. 
By a non-perturbative phase transition one can gauge transform these fields to 
the trivial fields \cite{ms1}-\cite{ms2}. Thus, in this picture the seed field
has been woken up from the non-Abelian vacuum in the electroweak  theory
by a phase transition.

In our estimate of the value of the magnetic field today we assumed 
statistical isotropy
and the validity of the standard MHD equations (\ref{mhd}) after the phase 
transition. The actual solution (\ref{ao}) of the electroweak theory 
ceases to be valid after some time, since the magnetic field
decreases due to the expansion of the universe and the non-linear 
dynamics of the MHD equations and hence after some time
reaches a value below the threshold $m_W^2/e$. The magnetic field then
becomes electroweak stable and there is  no longer 
any reason \cite{nkn} for the presence of the $W-$condensate, so these 
vector bosons can decay according to the standard theory.
The magnetic field is then subsequently
ruled entirely by the MHD equations (\ref{mhd}).

\end{document}